\documentstyle[prl,aps]{revtex}
\begin{document}
\def\Ca{\mbox{Ca$^{2+}$ }}
\def\IP3{\mbox{IP$_3$ }}
\draft
\input{psfig.sty}
\twocolumn[\hsize\textwidth\columnwidth\hsize\csname@twocolumnfalse\endcsname
\title{Discrete stochastic modeling of calcium channel dynamics}
\author{Markus B\"{a}r$^1$, Martin Falcke$^{2,3,}$\dag, Herbert Levine$^2$ and Lev S.~Tsimring$^4$}
\address{$^{1}$Max-Planck Institut f\"{u}r Physik komplexer Systeme,
Dresden, Germany\\
$^2$ Department of Physics, University of California at
San Diego, La Jolla, CA 92093-0319\\
$^3$ Department of Molecular Medicine, Institute of Biotechnology,
University of Texas Health Science Center at San Antonio, 15355 Lambda
Drive, San Antonio, TX  78245-3207\\
$^4$ Institute for Nonlinear Science, University of California at
San Diego, La Jolla, CA 92093-0402\\
}
\date{\today}

\maketitle

\begin{abstract}
We propose a simple  discrete stochastic model for calcium dynamics
in living cells. Specifically, the calcium concentration distribution
is assumed to give rise to a set of probabilities
for  the opening/closing of channels which release calcium
thereby changing those probabilities.  We study this model in one
dimension, analytically in the mean-field limit of large number of channels per
site $N$, and numerically for small $N$. As the number of channels
per site is increased, the transition from a non-propagating
region of activity to a propagating one changes in nature from one
described by directed percolation to that of deterministic depinning
in a spatially discrete system. Also, 
for a small number of channels a propagating calcium wave can leave behind
a novel fluctuation-driven state,  in a parameter range 
where the limiting deterministic model exhibits only single pulse propagation. 
\end{abstract}

\narrowtext
\vskip1pc]

\draft

It has become clear that the intracellular nonlinear dynamics of
calcium plays a crucial role in many biological processes \cite{Berridge98}. The
nonlinearity of this problem is due to the fact that there exist calcium
stores inside the cell which can be released via the opening of
channels which themselves have calcium-dependent kinetics. Typically,
these processes are modeled using a set of coupled equations for the
calcium concentration (the diffusion equation with sources and sinks)
and for the relevant channels; the latter is often described by a
rate equation for the fraction of open channels
per unit of area. More elaborate models take into account the discrete nature
of these channels, their spatial clustering, and fluctuations in the 
process of their opening and closing \cite{KeizerSmith98,Keizer98}. 

In this paper, we will propose and analyze a set of models which operate just
with the channel dynamics alone. The justification for this is that
the calcium field equilibrates quickly, with a diffusion time of
perhaps 0.1s, as compared to the channel transition times, perhaps
on the order of 1s for activation of a subunit to
several seconds for its deactivation. One can then imagine
solving for the  quasi-stationary calcium concentration and thereafter
using it to determine the conditional probabilities of channel opening or 
closing. In a subsequent paper\cite{FLT}, we will show how this can be done in
detail starting from a specific fully-coupled model (the DeYoung-Keizer-model
\cite{DeYoung,Keizer1}); here, we will make reasonable assumptions for these
probabilities and study the resulting stochastic model in a one dimensional
geometry.

For specificity, we will focus on systems that have \IP3 (inositol 1,4,5-
trisphosphate) channels. Each of these channels consists
of a number of subunits. Here we assume that $h$ subunits have to be 
activated for the channel to be open; experiments indicate
that $h=3$ \cite{Bezprozvanny}. A subunit is activated when \IP3 
ion is bound to its corresponding domain and \Ca is bound to its activating 
domain and {\em not} bound to its inhibiting site. The characteristic time 
of binding and unbinding of \IP3 is typically so fast (more than 20 times 
faster than other binding steps \cite{DeYoung}), that we can assume local 
balance of active/passive channels maintained at all times. Furthermore, 
we assume that the channels are spatially organized into clusters 
\cite{Parker1,Parker2}, with a fixed number of channels $N$ per cluster and 
a fixed inter-cluster distance. 

Our model is as follows.
We introduce two stochastic variables for each channel cluster: $n_i$,
the number of activated subunits, and $m_i$, the number of inhibited
subunits. At every time step, the
number of activated subunits $n_i$ at site $i$ is changed due to three
stochastic  processes; activation of additional subunits by binding
available \Ca to their activation domains, de-activation by unbinding
\Ca from active subunits, and inhibition by binding available \Ca to
their inhibition domains. We take these transition rates to depend 
on the number of open channels at site $i$, $c_i$, and on the number of
open channels at the nearest neighboring sites $i\pm 1$, $c_{i\pm 1}$. 
Similarly, there will be binding and unbinding to the inhibitory domain,
changing $m_i$. We denote by $p^\pm_{0(1)}$ the probability to
activate/inhibit a subunit per number of open channels at
the same site (0) or the neighboring site (1).  To compute the actual
probabilities, we need to multiply these by the number of open channels.
Here, we use the simple expedient of taking this to equal $n_i^h/hN_s^{h-1}$
where the total number of subunits $N_s=hN$;
this is easily shown to be the expected number of open channels for
large enough $N$. This approach allows us to avoid keeping explicit
account of each of the independent subunits. Also, we let 
$p^\pm_d$ be the deactivation and deinhibition probabilities 
which are $c$ independent.

Let us define the total probabilities $p^\pm=p_0^\pm+2p_1^\pm$
and the ``diffusion constant'' $\alpha=p_1^\pm/(p_0^\pm+2p_1^\pm)$.
We also denote $C_i(t)=(1-2\alpha) c_i(t)+\alpha
c_{i-1}(t) + \alpha c_{i+1}(t)$, which mimics the amount of
calcium at site $i$ due to open channels at sites $i,\ i\pm1$.
Our model explicitly consists of the following coupled stochastic
processes. $n_i$ is updated
\begin{equation}
n_i(t+\Delta t)=n_i(t)+\Delta^+_n -\Delta^-_n - \delta_+
\end{equation}
where $\Delta^+_n$ is a random integer number drawn from the binomial
distribution $B(\Delta^+_n,N_s-n_i(t)-m_i(t),p^+C_i(t))$, $\Delta^-_n$ is
drawn from $B(\Delta_n^-,n_i(t),p^-C_i(t))$, and $\delta^+_n$ is drawn
from  $B(\delta^+_n,n_i(t),p_d^+)$. The equation for $m_i$ reads 
\begin{equation}
m_i(t+\Delta t)=m_i(t)+\Delta^+_m - \delta^+_m
\end{equation}
where $\Delta^+_m$ is drawn from  $B(\Delta^+_m,N_s-m_i(t),p^-C_i(t))$, and
$\delta^+_m$ is drawn from $B(\delta^+_m,m_i(t),p_d^-)$. We do not allow
for transitions from the inhibited state to the activated state. In all
these formulas, $B(x,y,p) \equiv {y \choose x} 
p^x (1-p)^{y-x}$. Note that the probability that \IP3 is bound is included by
rescaling the number of subunits.

\begin{figure}
\centerline{ \psfig{figure=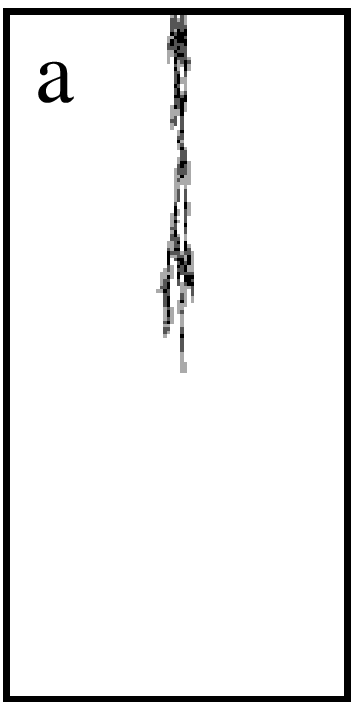,width=1.in} 
\psfig{figure=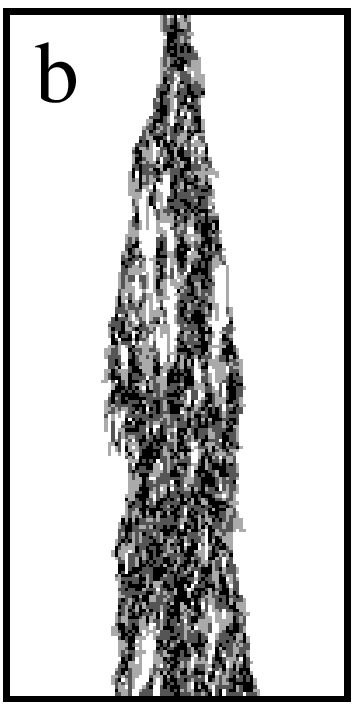,width=1.in} 
\psfig{figure=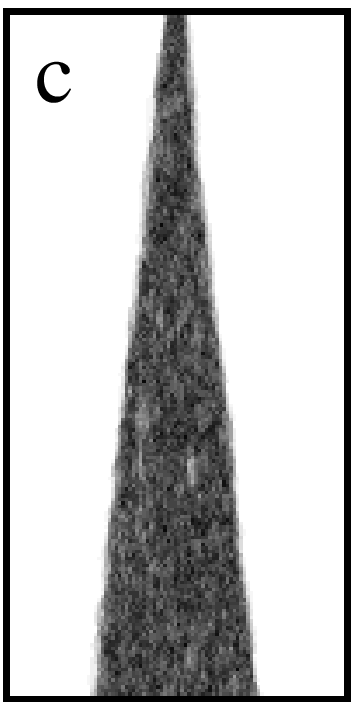,width=1.in} }
\caption{Evolution of an initial seed of 5 clusters of open channels in the 
middle of the lattice for $p^+N_s/h=1$, $p^+_d=0.2$, $h=3$, and $N_s=3$, 
$\alpha=0.1$ {\protect\em (a)}, $N_s=3$, $\alpha=0.25$ {\protect\em (b)}, and  
$N_s=30$, 
$\alpha=0.25$ {\protect\em (c)}. The horizontal axis is spatial coordinate (100
sites), and the vertical axis is time (1000 iterations).}
\label{sim1}
\end{figure}

As a first step, we consider a simplified version of the channel
dynamics with the inhibition process excluded (all $p^-$=0), i.e. a
subunit is activated whenever \Ca is attached to its activating site.
Thus we take $m_i=0$, and arrive at the one-variable model for the
number of activated subunits $n_i$.  Let us first focus on fairly small $N_s$.
Examples of the stochastic 
dynamics for several values of parameters are shown in Figure
\ref{sim1}.  At small $\alpha$, an initial seed almost always ultimately
dies giving rise to so-called abortive calcium waves. At larger values
of $\alpha$ the region of activated channels typically expands at a
finite rate.  This transition mirrors what has been seen in many 
experimental systems \cite{Parker2}.

\begin{figure}
\centerline{\psfig{figure=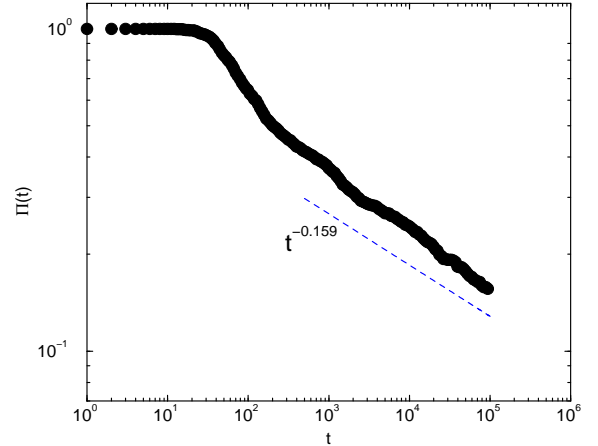,width=3.2in}}
\caption{The distribution of survival time for the stochastic
model with $h=3$, $\gamma=0.1$, $N_s=10$, and $\alpha=\alpha_c=0.359$.
Dashed line indicates the power-law scaling $\propto t^{-0.159}$.}
\label{survival_h3}
\end{figure}

As is well known for statistical models such as the
contact process\cite{Harris}, the critical value of $\alpha$ can be accurately determined
by computing the distribution of survival 
times $\Pi(t)$ for the activation process started from a single active site. 
For $\alpha<\alpha_c$, the distribution falls exponentially at large $t$ 
as the wave of activation eventually dies out. On the contrary, at 
$\alpha>\alpha_c$, $\Pi(t)$ asymptotically reaches a constant value 
$\Pi_\infty$, since a non-zero fraction of runs produce ever-expanding
active regions. At $\alpha=\alpha_c$, the distribution function exhibits
a power-law asymptotic behavior with the slope determined by the universality
class of the underlying stochastic process.  Our data (not shown)
indicate that $\alpha_c$ is inversely proportional to the number of subunits per
site $N_s$. We have checked that our
data is in  the {\em directed percolation} (DP)\cite{DP} class. For example, 
in Fig.  \ref{survival_h3} we show $\Pi (t)$ of a cluster of open
channels at the critical value of $\alpha_c$ for $h=3$, $N_s=10$ and $\gamma=0.1$.
The power-law dependence is consistent with DP prediction of $\Pi
(t) \propto t^{-0.159}$.  This is perhaps not too
surprising. According to the Janssen-Grassberger
DP conjecture\cite{JanGras}, any spatio-temporal stochastic process with
short range interactions, fluctuating active phase and unique
non-fluctuating (absorbing) state, single order parameter and no
additional symmetries, should belong to the DP class. This result does
open up the exciting possibility that intracellular calcium dynamics could
be an experimental realization of the DP process.

Figure \ref{sim1}(c) shows the opposite limit where the dynamics becomes
almost deterministic. If we take
$N_s\to\infty$ and fix $pN_s/h\to P$, we can use a mean-field
description in terms of the fraction of activated
subunits $\rho_i=n_i/N_s$,
\begin{equation}
\dot{\rho_i}=((1-2\alpha)\rho_i^h+\alpha\rho_{i-1}^h + \alpha\rho_{i+1}^h)
(1-\rho_i) -\gamma\rho_i.
\label{mf}
\end{equation}
and where we rescaled time $t'=Pt/\Delta t$ and introduced $\gamma=p_d/P$.
For all $h\geq 2$, if $\gamma<\gamma_{cr}$ Eq.(\ref{mf}) the system
possesses two stable uniform solutions, $\rho=0$ and 
$\rho=\rho_0$ and one unstable solution $\rho_u$, where $\rho_{0,u}$ are 
real roots of the algebraic equation $\rho^{h-1}(1-\rho)=\gamma$. The
front is a solution connecting these two stable fixed points; it is easy
to show that this front has a unique propagation velocity.

\begin{figure}
\centerline{\psfig{figure=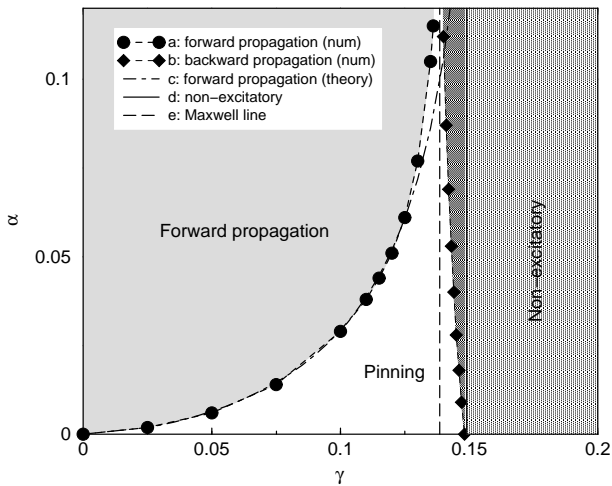,width=3.2in}}
\caption{Phase diagram of the mean field equation (\protect\ref{mf}) for $h=3$: {\em a} 
bifurcation line separating forward-propagating fronts from pinning region; 
{\em b} same for backward propagating fronts; {\protect\em c} small-$\alpha$ 
approximation of pinning line; {\em d} line 
$\gamma=4/27$ separating the region of non-existence of the excited state;
{\em e} Maxwell line $\gamma=0.138...$ separating forward and backward front 
propagation in the continuum limit}
\label{phdiag}
\end{figure}

For small $\alpha$, the discreteness of our spatial lattice
causes  the front to become pinned, as
the probability of activating subunits at the neighboring site 
$O(\alpha\rho_0^h)$ becomes smaller than the threshold value for excitation 
probability $O(\rho_u)$. The stationary front solution is described by the 
recurrence relation,
\begin{equation}
(1-2\alpha)\rho_i^h+\alpha\rho_{i-1}^h + \alpha\rho_{i+1}^h=
\frac{\gamma\rho_i}{1-\rho_i}
\label{stfront}
\end{equation}
The bifurcation line which separates pinned and 
moving fronts, can be found in the limit of small $\alpha$ by using the
ideas of ref.\cite{mitkov}. Indeed, in this 
limit, the values of $\rho_i$ quickly (as $\alpha^i$) approach 0 and 
$\rho_0$ away from the front at $i\to\pm\infty$, respectively. We can thus
replace $\rho_i$ by $\rho_0$ and 
$0$ everywhere to the left and to the right of the front position except for 
$\rho_\pm$ at the two sites nearest to the front, $i-1$ and $i+1$.
Solving the resulting set of two algebraic equations up to 
$\alpha^2$, one can obtain the values of $\rho_\pm$. At any $\gamma$, 
there is a critical value of $\alpha_m$ at which the real solution 
$\rho_\pm$ vanishes. The family of these values $\alpha_m$ forms the 
bifurcation line for front pinning in $(\gamma,\alpha)$ plane. At large 
$\alpha$, discreteness of the mean field model (\ref{mf}) becomes 
insignificant, and (\ref{mf}) can be replaced by its continuum limit
\begin{equation}
\partial_t\rho=(\rho^h-\alpha\partial_{x}^2\rho^h)(1-\rho) -\gamma\rho.
\label{cont}
\end{equation} which
of course has no front pinning. Instead, $\alpha$ can be scaled
out and there is specific value of $\gamma$
at which the system goes from forward to backward propagating fronts.
Figure \ref{phdiag} shows the phase diagram of the mean field equation 
(\ref{mf}) for $h=3$. All the data
(except possibly at the non-generic case $\gamma=0$) are consistent with expected\cite{mitkov} 
$(\alpha-\alpha_m)^{1/2}$ scaling.

\begin{figure}
\centerline{\psfig{figure=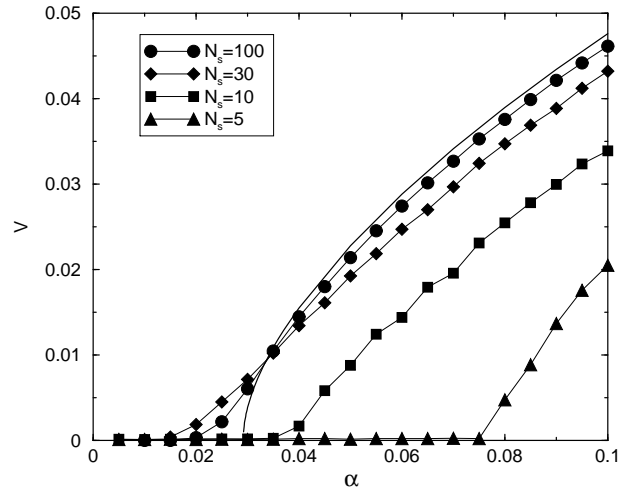,width=3.2in}}
\caption{The average front speed as a function of $\alpha$ for
stochastic model at $h=3$, $\gamma=0.1$, $p^+N_s/h=1$, $p^+_d=0.1$, 
and different values of $N_s$. Solid line indicates the mean-field 
limit $N_s\to\infty$.}
\label{vel_h3}
\end{figure}

How does one get from DP behavior to deterministic pinning/depinning?
To investigate this issue, we have performed simulations for the
front speed as a function $\alpha$ at various finite values of $N_s$;
with the results given in Fig. \ref{vel_h3}. At large $N_s$, the velocity 
approaches the mean field prediction as long as $\alpha>\alpha_m$. 
Close to  critical value
$\alpha_m$, the velocity deviates from the mean-field dependence 
$V\propto(\alpha-\alpha_m)^{1/2}$ because of thermally activated
``creep''; fluctuations let the front to overcome potential
barriers associated with finite site separation, and lead to exponentially 
slow front propagation (see, e.g., \cite{ampt}). Directed percolation regime is
not observed at large $N_s$ since the DP critical value $\alpha_c$ is less
than $\alpha_m$. At smaller $N_s$, the relative
magnitude of the fluctuations grows, and the DP threshold value $\alpha_c$
exceeds $\alpha_m$. Now, the front pinning is determined by
fluctuations rather than discreteness, and the critical state exhibits the
properties of directed percolation. 

\begin{figure}
\centerline{\psfig{figure=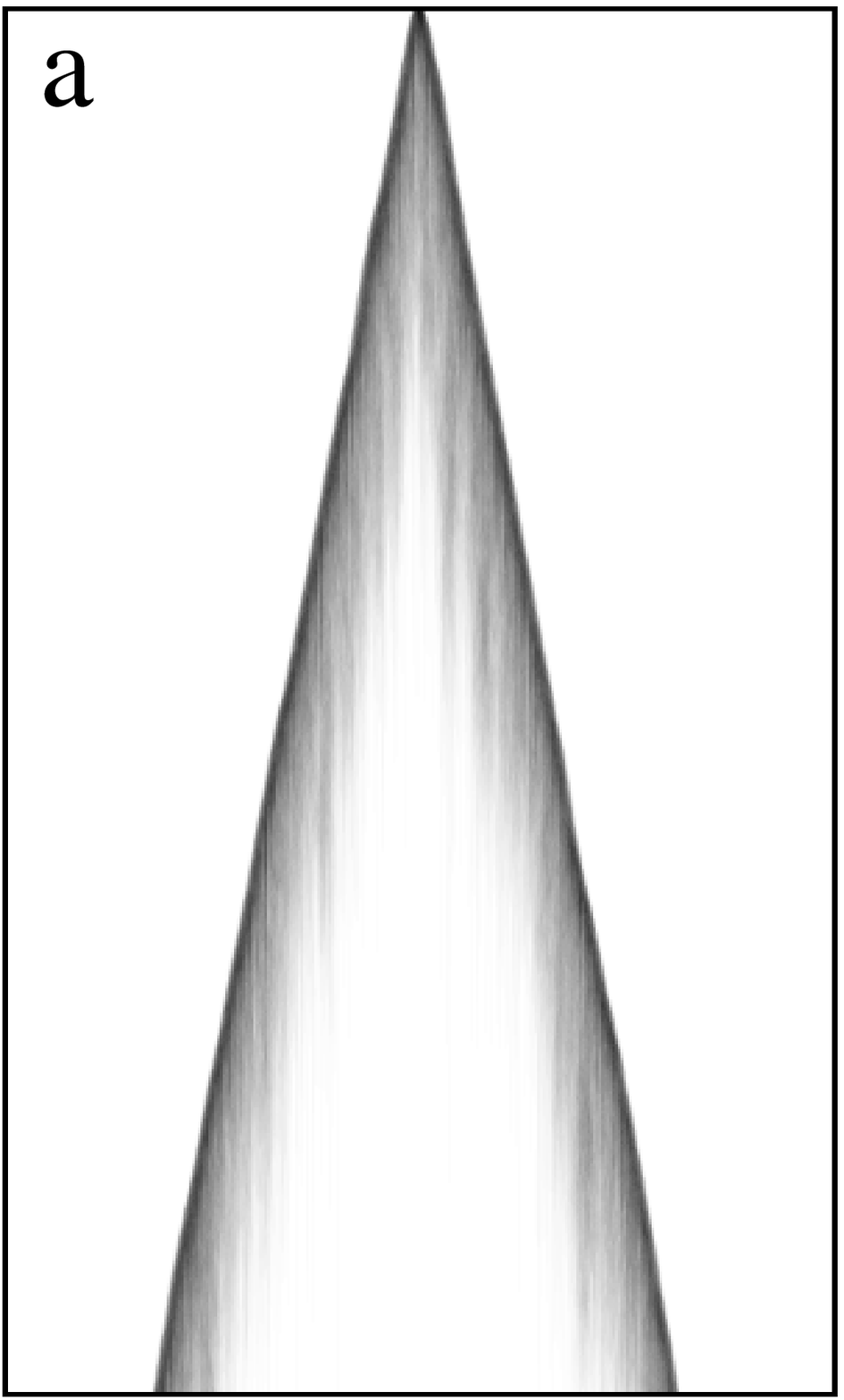,width=1.5in}
\psfig{figure=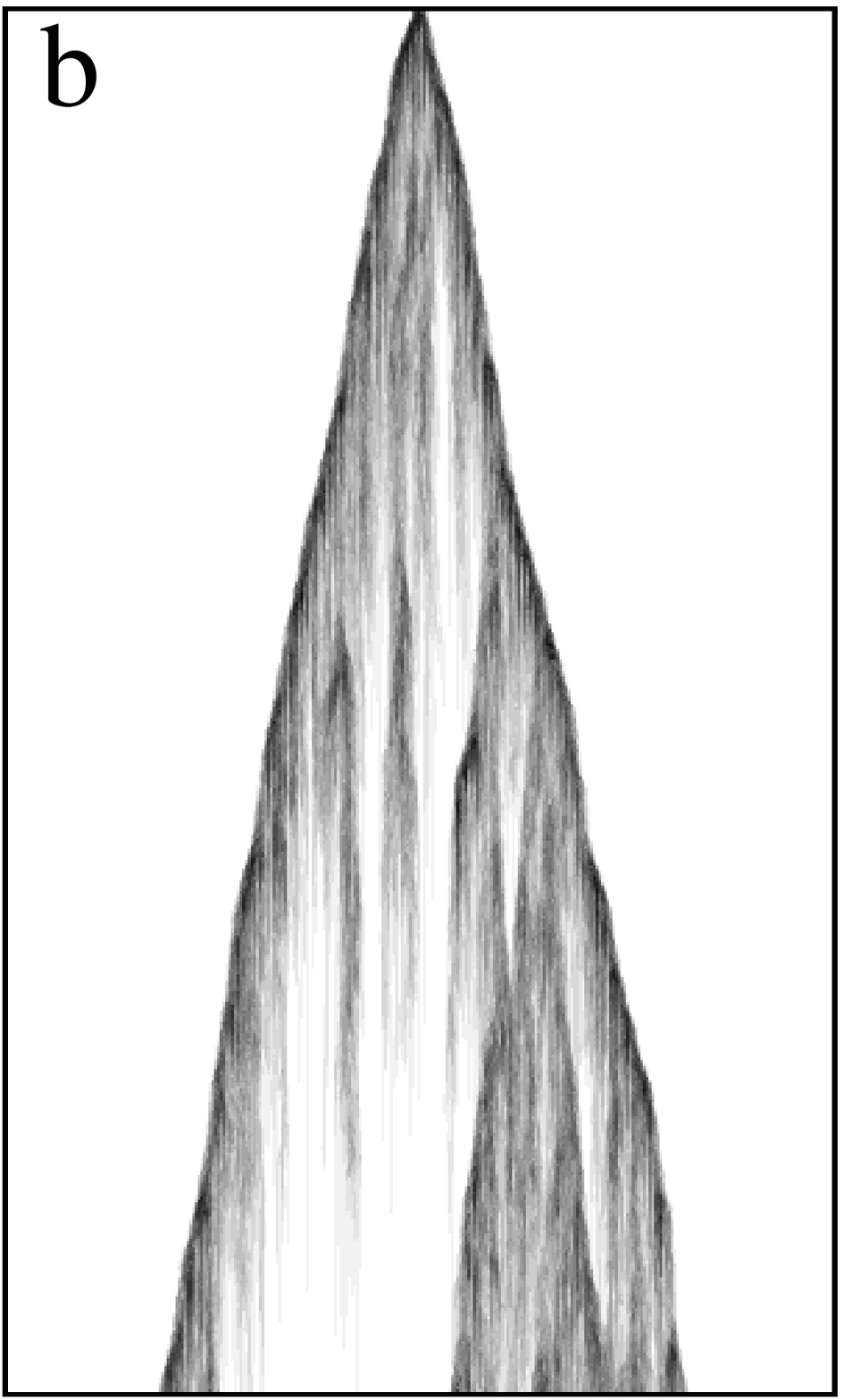,width=1.5in}}
\caption{Space-time evolution initiated by opening channels at a single
cluster in the middle of the lattice of 300 sites for full 
activation/inhibition model with $p^+=1$, $p^+_d=0.04$, $p^-=0.1$, 
$p_d^-=0.12$, $h=3$, $\alpha=0.7$, and $N_s=200$ (a) and  $N_s=20$ 
(b),  500 iterations.}
\label{sim2}
\end{figure}

Now we return to the full two-variable stochastic model which describes
both activation and inhibition. Since the probability of \Ca
binding to the inhibition domain is typically much smaller
than those for the activation domain, the inhibitor dynamics is 
slow. In the mean-field limit $N_s\to\infty$, this model is similar to the
FitzHugh-Nagumo model often used to describe waves propagating in
excitable systems. One therefore expects that for a certain range of 
binding/unbinding probabilities, the model gives rise to pulse
propagation; that is, once the wave passes, the system goes into a
state dominated by inhibition from which it slowly recovers as the
inhibitory domains slowly unbind. This is indeed what we find
for large enough $N_s$, as shown in Fig. \ref{sim2}(a). Behind the
pair of outgoing pulses, the
channels stay refractory for a certain time $O(1/p^-)$ and then return to
the quiescent state.

However, we find that having only a modest number of channels $N$ 
leads to fluctuations which strongly affect the
spatio-temporal behavior of the model.  In fact, a new 
dynamical state is formed behind 
the outgoing fronts, a state which remains active at all 
subsequent times (see Fig.\ref{sim2},b). This state is catalyzed by
backfiring, i.e. the creation of oppositely propagating waves behind
a moving front. In the deterministic limit of our model, this cannot occur as the
system is completely refractory once the front has passed. At finite $N$ 
however, propagation of the front does not lead to the activation and 
subsequent inhibition of all the channels. Instead, a finite number of 
these remain inactivated, providing a supply of active elements that 
can still support wave propagation. There exist more complicated deterministic
models\cite{zimmerman}, such as one proposed for $CO$ oxidation on single crystal
surfaces\cite{Ertl}, which also appear to have 
pulse-induced backfiring. There, however,
this effect is due to the loss of pulse stability which occurs due
to the rather complex non-linear dynamics of the inhibitory field. Here,
it is the fluctuations which allow for this phenomenon.

We have checked that this backfiring-induced state 
occurs as well in more realistic and more complex models which solve 
for the calcium concentration together with the channel dynamics. Again,
the mechanism appears to be the lack of complete inhibition in
the wake of the propagating pulse. Hence, our result that one
should find this behavior in intracellular calcium dynamics
is not an artifact of any of the simplifying assumptions used here. Also, this 
state persists when the model is studied in higher dimensions. A study of the 
exact nature of the transition 
to backfiring and a comparison of the deterministic versus
stochastic pathways to its existence will be undertaken in future
work\cite{FLT}.

In summary, we proposed and studied a simple discrete model of 
calcium channel dynamics based on the assumption that calcium diffusion
time is much smaller than the characteristic times of \Ca binding/unbinding.
This model demonstrates familiar properties of deterministic 
reaction-diffusion systems in the limit $N\to\infty$ when fluctuations
are small. For small $N$, we observed a transition to a directed
percolation regime, in agreement with the general DP
conjecture\cite{JanGras}. For the full model including
inhibition,  we found at small $N$ a novel persistent fluctuation driven state 
which emerges behind a front of outgoing activation; this
occurs in a parameter regime where the corresponding
deterministic system  exhibits only single outgoing pulses.

The authors thank H. Hinrichsen, M.Or-Guil and I. Mitkov for helpful
discussions.  LST thanks Max Planck Institut f\"{u}r Physik komplexer
Systeme, Dresden, Germany for hospitality. LST was supported in part by
the Engineering Research Program of the Office of Basic Energy Sciences
at the US Department of Energy under grants No. DE-FG03-95ER14516 and
DE-FG03-96ER14592. HL was supported in part by US NSF under grant
DMR98-5735. M.F. was supported in part by DFG grant Fa350/2-1.

\end{document}